\documentclass[aps,prd,preprint,groupedaddress,showpacs]{revtex4}
\usepackage{graphicx}

\newcommand{\delete}[1]{}

\newcommand{\be}{\begin{equation}}
\newcommand{\ee}{\end{equation}}

\def\beq{\begin{equation}}
\def\eeq{\end{equation}}
\def\bea{\begin{eqnarray}}
\def\eea{\end{eqnarray}}
\def\ba{\begin{array}}
\def\ea{\end{array}}

\begin{document}

\preprint{Stanford-ITP-03-17}

\title{Probing Sub-Micron Forces by Interferometry \\ of Bose-Einstein Condensed Atoms}


\author{Savas Dimopoulos}
\email[]{savas@stanford.edu}
\author{Andrew A. Geraci}
\email[]{aageraci@stanford.edu}
\affiliation{Department of Physics, Stanford University, Stanford,
CA 94305}


\date{\today}

\begin{abstract}
We propose a technique, using interferometry of Bose-Einstein
condensed alkali atoms, for the detection of sub-micron-range
forces. It may extend present searches at 1 micron by 6 to 9
orders of magnitude, deep into the theoretically interesting
regime of 1000 times gravity. We give several examples of both
four-dimensional particles (moduli), as well as higher-dimensional
particles -- vectors and scalars in a large bulk-- that could
mediate forces accessible by this technique.
\end{abstract}


\maketitle

\section{Introduction}

Some recent theoretical ideas point to the possibility of new
physics, related to gravity, at the sub-millimeter regime. One is
the preponderance of light gravitationally coupled moduli
suggested by string theory \cite{sg,add97}. Another is the
possibility of large sub-mm-size dimensions and the particles
residing inside its bulk \cite{add1,iadd,add3}. Yet another is
suggested by the magnitude of the vacuum energy \cite{raman97}.

These ideas motivated some heroic experiments
\cite{stanford,pricenature,price,adelberger,pricereview} that
have, in the last seven years, extended the search for such forces
from mm down to $\sim 20$ microns. These experiments involve
measuring the force between two macroscopic but small objects.  A
fundamental obstacle in searching at much smaller distances is
that the size of these objects must be reduced and therefore the
expected signal force decreases; at the same time, the
electrostatic background Van-der-Waals force increases.

In this paper we suggest a possible way around this obstacle by
considering the interaction of a macroscopic system with a pure
quantum mechanical system consisting of a Bose-Einstein
condensate. The latter has a significant advantage relative to a
macroscopic system:  its de Broglie phase can be measured very
precisely. In addition, it can be well controlled and manipulated
and its electromagnetic interaction with its environment is well
understood -- both theoretically and experimentally. Because of
these advantages, the technique that we propose may extend current
bounds at 1 micron by 6 to 9 orders of magnitude, and be sensitive
to forces as small as 1000 times gravity.  The approach we
describe thus explores a region of parameter space that is
complementary to the super-micron reach of upcoming
micro-cantilever and torsion balance experiments.

In section 2 we update the analysis of macroscopic forces below 10
microns in theories with light moduli.  In section 3  we consider
new forces from bulk gauge fields or scalars in large extra
dimensions, taking baryon number as an example. In section 4 we
propose our experimental technique and estimate some of the
important backgrounds. We conclude with section 5.

\section{Forces from Light Moduli}

In string theory the parameters of the standard model depend on
fields, called moduli, whose values determine the geometry of the
extra dimensions. Moduli couple with gravitational strength and
typically remain massless until supersymmetry is broken. So, they
get a mass proportional to  $\sim F/M_{\rm PL}$, where $F$ is the
scale where supersymmetry breaking originates. In theories with
gravity-mediated supersymmetry breaking, $F$ is $(10^8~{\rm
TeV})^2$ and the moduli have microscopic Compton wavelengths.
However, as pointed out in reference \cite{sg}, in theories of
gauge-mediated supersymmetry breaking, $F$ can be as small as
$(10~{\rm TeV})^2$ and moduli can have macroscopic Compton
wavelengths and mediate macroscopic forces of gravitational
strength. The range and magnitude of these forces, for a variety
of moduli, were first estimated in reference \cite{sg} for
$\sqrt{F}$ in the range of 10 TeV to 100 TeV. At that time it
seemed pointless to consider larger values of F, since they lead
to moduli Compton wavelengths which were thought to be
inaccessible to macroscopic-force experiments. In this paper we
extend the scale of $\sqrt{F}$ up to $2000$ TeV, which in turn
considerably extends the predicted parameter space for
moduli-dependent forces.

The upper limit for the value of $\sqrt{F}$ comes from cosmology:
In gauge-mediated supersymmetry breaking, the gravitino is the
lightest supersymmetric particle, with mass $\sim F/M_{\rm PL}$.
Although light, its mass still must not exceed $1$ keV to avoid
over-closing the universe \cite{giudicereview}. This in turn
provides an upper limit of $2000$ TeV on $\sqrt{F}$.

We focus on the three classes of moduli studied in reference
\cite{sg} which couple directly to ordinary matter: the dilaton,
the gauge moduli and the Yukawa moduli.  Moduli-dependent forces
can occupy a substantially larger region of parameter space than
previously indicated.  Although much recent experimental progress
has been made in the search for new sub-millimeter forces
\cite{pricereview},\cite{fischbach} there is ample potentially
interesting parameter space awaiting further exploration.

\begin{figure}
\includegraphics[width=1.0 \columnwidth]{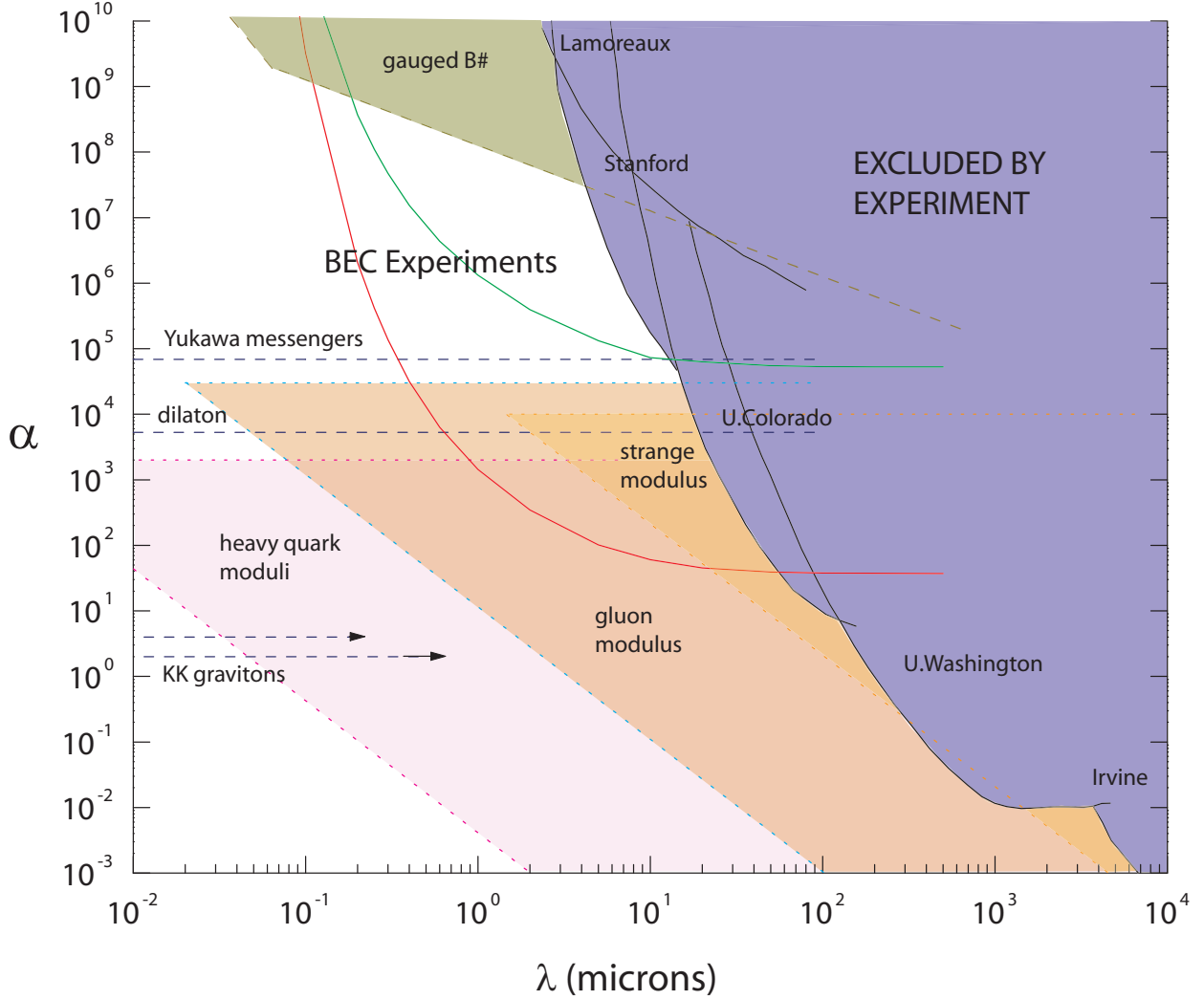}
\caption{Experimental bounds and theoretical expectations on new
forces from potentials of the form $V(r) = -G_N \frac{m_1 m_2}{r}
(1 + \alpha e^{-r/\lambda})$ below 1 cm.  The projected reach of
the first-round BEC experiments is shown as a solid green line.
The solid-red line indicates the reach with an improved
sensitivity of $10^{-7}$Hz. Experimental data are from references
\cite{stanford,pricenature,adelberger,irvine,lamoreaux}. The shown
theoretical expectations are discussed in the text.} \label{submm}
\end{figure}
\begin{figure}
\includegraphics[width=1.0 \columnwidth]{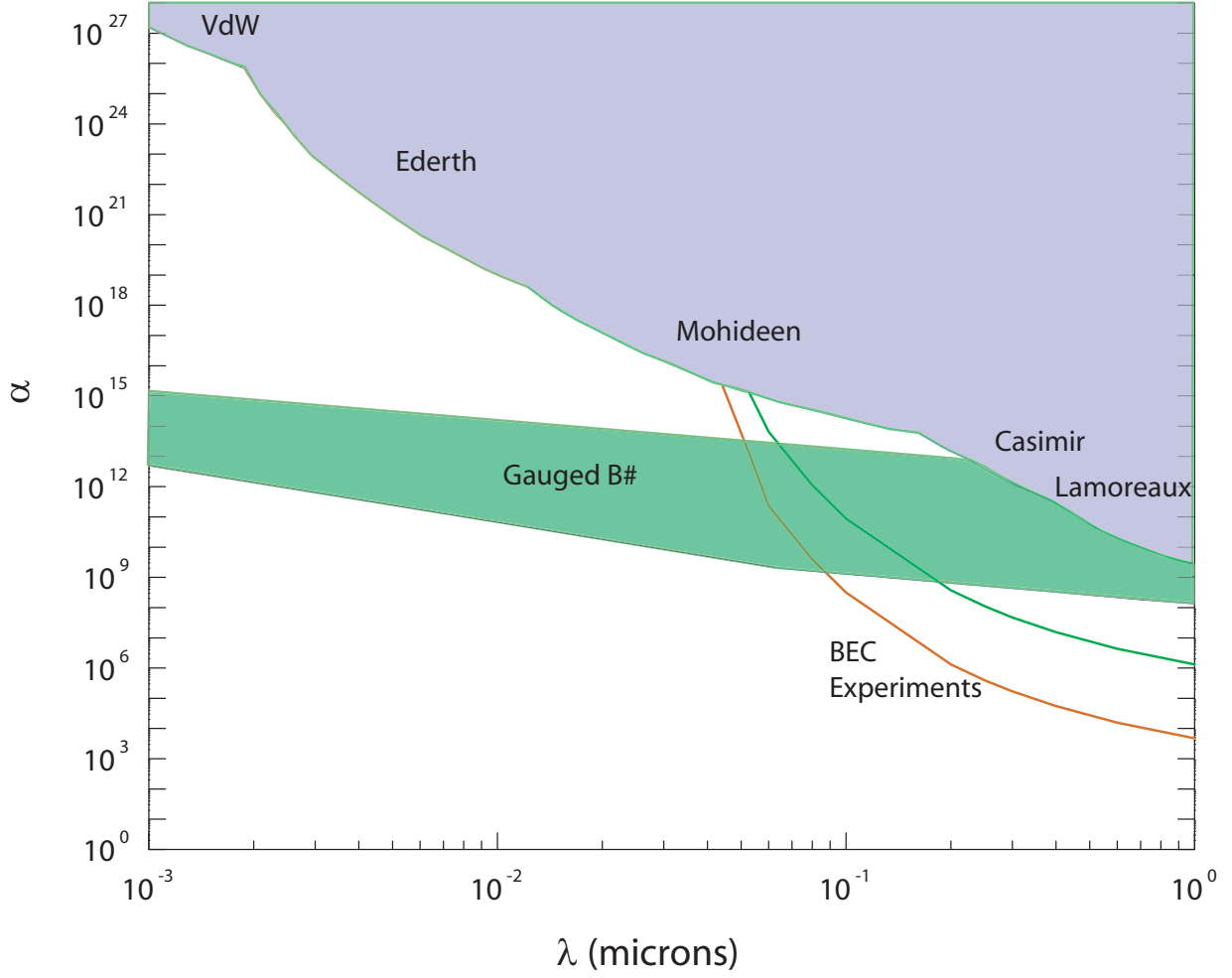}
\caption{Experimental bounds on new forces from potentials of the
form $V(r) = -G_N \frac{m_1 m_2}{r} (1 + \alpha e^{-r/\lambda})$
below 1 micron.  The projected reach of the first-round BEC
experiments is shown as a solid green line.  The solid-red line
indicates the reach with an improved sensitivity of $10^{-7}$Hz.
Experimental data are adapted from a figure in reference
\cite{fischbach}} \label{subum}
\end{figure}

\subsection{Gluon modulus}
We consider here a field $\phi$ that couples only to the standard
model gluons.  The effective coupling is given by \cite{sg} \beq
{\mathcal L}=\frac{\lambda_g}{8\pi^2}\frac{\phi}{M}G_{\mu
\nu}^aG^{a \mu \nu}~, \label{glue} \eeq where $\lambda_g$ is an
undetermined coupling constant, $M$ is expected to be of the order
of the string scale $5\times 10^{17}$ GeV, and the suppression
factor $8\pi^2$ accounts for the gauge coupling depending on
moduli only at higher-order. In this way, the coupling strength is
weaker than that of the dilaton discussed in the following
subsection.

Considering the contribution to its mass coming from the
interaction in eq.~(\ref{glue}), the Compton wavelength of the
field is \cite{sg} \beq \lambda_\phi = 8\times 10^{-4}~{\rm
m}~\lambda_g^{-1} ~\left(\frac{M}{5\times 10^{17}~ {\rm
GeV}}\right)~\frac{(100~{\rm TeV})^2}{F}~(kN)^{-1/2}~. \eeq Here
$F$ is the fermion-scalar messenger mass-squared splitting, $k$ is
a loop-integral factor of order 1, and $N$ is the number of
messenger multiplets.  $N<4$ is required so that all gauge
couplings remain perturbative below the GUT scale, and a bound of
$\sqrt{F}> 30 ~\rm{TeV}/\sqrt{N}$ can be imposed from constraints
on the right-handed selectron mass and a consistency condition
that the messengers have non-negative mass-squared \cite{sg}.

The coupling of $\phi$ to the nucleon $N$ can be expressed by \beq
{\mathcal L}={\mathcal G}_{\phi}\frac{m_N}{M_{\rm PL}} \phi {\bar
\psi}_N \psi_N~, \label{couppp} \eeq where ${\mathcal G}_\phi$ is
given by \beq {\mathcal G}_\phi
=\frac{\lambda_g}{8\pi^2}\frac{M_{\rm PL}}{M} \frac{\langle N |
G_{\mu \nu}^aG^{a \mu \nu} |N\rangle }{m_N} \simeq -6~\lambda_g
~\left(\frac{5\times 10^{17}~{\rm GeV}}{M} \right)~. \eeq The
interaction in eq.~(\ref{couppp}) yields a potential for $\phi$
exchange between particles of mass $m_1$ and $m_2$ at a distance
$r$: \beq V(r)=G_N m_1m_2{\mathcal G}_\phi^2\int
\frac{d^3k}{(2\pi)^3}\frac{e^{i\vec{k} \cdot
\vec{r}}}{\vec{k}^2+m_\phi^2}~. \label{vecpot} \eeq When added to
gravity, eq.~(\ref{vecpot}) describes an additional attractive
force: \beq V(r) = - G_N \frac{m_1m_2}{r}\left(
1+\alpha~e^{-r/\lambda_{\phi}}\right)~, \eeq where $\alpha$ is
given by ${\mathcal G}_{\phi}^2/{4 \pi}$.  In reference \cite{sg},
a range of $\lambda_{\phi_g}$ and $\alpha$ was obtained by taking
$\sqrt{k}N=1$ and by varying $\sqrt{F}$ between 30 and 100 TeV and
$\lambda_g^{-1}\times M/(5\times 10^{17}~{\rm GeV})$ between
$10^{-2}$ and $10^2$.  We here expand the range for the SUSY scale
$\sqrt{F}$ up to the limit of $2000$ TeV imposed by the gravitino
problem and show the results in Figure \ref{submm}.

\subsection{Dilaton}

The dilaton couples to nucleons with a strength of about 80 times
gravity \cite{ven,sg,kw}, leading to an inter-nucleon force of
about 6400 times gravity. Since it couples to all fields in the
theory, it is expected to receive a mass $\sim F/M$ from its
strong coupling to the primordial supersymmetry-breaking sector.
This would make its Compton wavelength less than $10^{-2}~\mu{\rm
m}~(100~{\rm TeV})^2/F$, which is too short to be experimentally
observed. However, as Damour and Polyakov speculate \cite{dp},
since the dilaton potential is related to the cosmological
constant, the yet unknown mechanism accounting for the smallness
of the cosmological constant may also make the dilaton light.
Since there is no good theory of the dilaton mass, the line
labelled "dilaton" in Figure 1 should be terminated a point
determined by experiment.

\subsection{Yukawa moduli}
The Yukawa couplings of the standard model could also depend on
moduli which are relatively unaffected by Planck scale physics,
but obtain a mass due to (low-scale) supersymmetry breaking
\cite{sg}. A Yukawa modulus $\phi$ may be coupled as follows to
up-type quarks and the Higgs boson $H$:

\beq {\mathcal L}=\lambda (\phi )q_L {\bar u}_R H_u + {\rm h.c.}
\label{int}~, \eeq along with analogous terms for down-type quarks
and charged leptons.  Here flavor indices have been suppressed and
we for simplicity assume one modulus per coupling.

Yukawa terms of the form in eq. (\ref{int}) contribute to an
effective potential for $\phi$:
\[
V(\phi) = \frac{8kN \alpha_s^2}{3(16\pi^2)^3} \lambda^\dagger
(\phi) \lambda(\phi) F^2 + V_0(\phi)
\]
where again $F$ is the fermion-scalar messenger mass-squared
splitting, $k$ is a loop-integral factor of order 1, $N$ is the
number of messenger multiplets, and $V_0$ describes any additional
unknown contribution to the potential not due to the operator in
eq. (\ref{int}).  The coupling can be expanded around its minimum
$\langle \phi \rangle \sim M$, where $M$ is of order string scale,
\[
\lambda (\phi) = \lambda^{(0)} + \lambda^{(1)} \frac{(\phi-\langle
\phi \rangle)}{M} + \frac{1}{2} \lambda^{(2)} \frac{(\phi-\langle
\phi \rangle )^2}{M^2} + ...
\]
and a lower bound on the modulus mass can be obtained from the
known term in
\[
m_\phi^2 = \frac{16kN \alpha_s^2}{3(16\pi^2)^3}(\lambda^{(1)2} +
\lambda^{(0)}\lambda^{(2)}) \frac{F^2}{M^2} + \frac{d^2 V_0
}{d\phi^2}|\phi= \langle \phi \rangle .
\]
For a particular flavor modulus, the coupling $\lambda^{(0)} = 2
m_q/\sqrt{2} v H_q$ where $v$ is the Higgs vacuum expectation
value of $246$ GeV, and $H_q$ equals $\sin{\beta}$ for up-type
quarks and $\cos{\beta}$ for down-type, where $\tan{\beta}$ is the
ratio of the Higgs' vacuum expectation values.  Using this result,
the Compton wavelength becomes
\begin{eqnarray*}
\lambda_\phi &=& 1050 \mu m \times \alpha_s^{-1}
\frac{F}{({\rm{100TeV}})^2} \frac{5 \times 10^{17} {\rm{GeV}}}{M}  \\
&\times& (\frac{{\rm{GeV}}}{m_q}) (\frac{H_q}{\frac{1}{\sqrt{2}}})
\left[(\frac{\lambda^{(1)2}}{\lambda^{(0)2}} +
\frac{\lambda^{(2)}}{\lambda^{(0)}})kN \right]^{-1/2}~.
\end{eqnarray*}
We arrive at an expression larger than reported previously in
reference \cite{sg} due to a corrected numerical factor.

The long-range force potential can be determined by relating the
scalar $\phi$-quark coupling of eq.~(\ref{int}) to the scalar
$\phi$-nucleon coupling.  The fields $\phi$ corresponding to up
and down quarks can generally have different couplings to the
proton and the neutron, leading to small violations of the
equivalence principle.  The scalar coupling of the field $\phi$ to
the nucleon $N$ is again expressed in terms of eq. (\ref{couppp})
where \beq {\mathcal G}_\phi =
\frac{\lambda^{(1)}}{\lambda^{(0)}}~\frac{M_{\rm PL}}{M}~
\frac{\langle N | m_q {\bar q} q |N\rangle }{m_N}~. \eeq

The results for the Compton wavelengths and the strengths of the
moduli forces relative to gravity are plotted in Figure
\ref{submm}.  The areas are obtained by taking $kN=1$,
$\tan\beta=1$, $\lambda^{(2)} =0$, by varying $\sqrt{F}$ between
30 and 2000 TeV, and by varying $\lambda^{(0)}
/\lambda^{(1)}\times M/(5\times 10^{17}{\rm GeV})$ between
$10^{-2}$ and $10^2$.

\section{Forces from particles in large extra dimensions}

In theories with large extra spatial dimensions the fundamental
scale $M_*$ and the observed four-dimensional Planck scale $M_4 =
2.43 \times 10^{18}$ GeV are related by
\[
M_4^2 = M_*^{{n+2}}V_{n}
\]
where $n$ is the number of extra dimensions and $V_n$ is their
volume.  In this framework, the standard model gauge and matter
content is confined to a 4-dimensional submanifold, and the
graviton can propagate in all $4+n$ dimensions. The scenario
provides an alternative solution to the gauge hierarchy problem
\cite{add1,iadd,add3}, as the fundamental scale can be of order
TeV. Such a paradigm also predicts a modification to Newton's law
of gravitation at distances nearby and below the length-scale of
compactification.
\subsection{Gravitons in the Bulk}
The modification of Newton's law of gravitation in theories with
large extra spatial dimensions has been studied in some detail
\cite{kands,fandl}.  For equal-size extra dimensions and toroidal
compactification, the volume satisfies $V_n = (2\pi R)^n = L^n $,
and we have \cite{add3}
\[
R_n = 2 \times 10^{31/n-16} {\rm{mm}} \times (\frac{1
{\rm{TeV}}}{M_{4+n}})^{1+2/n}
\]
For the case of two equal extra dimensions (n=2), the radius of
compactification is of order 1-millimeter for $M_{4+n} \sim $ TeV.
However for this case, astrophysical bounds require the
fundamental scale $M_{4+n}$ to be pushed above $1600$ TeV
\cite{raffelt}. At this scale the two equal-size radii are only
$2.4~\rm{\AA}$.  Requiring such a high energy scale in turn
necessitates more fine tuning for the framework to address the
hierarchy problem. For the case of 3 extra dimensions $M_{4+n}$
must exceed $60~\rm{TeV}$, and the radius becomes $\sim .05~
\rm{\AA}$ for equal-size dimensions. These limits are derived from
Kaluza-Klein gravitons that would be gravitationally trapped and
remain as a halo surrounding neutron stars \cite{raffelt}. The
constraints come from neutron star heating via Kaluza-Klein
graviton decays. Somewhat weaker limits are also obtained from
EGRET gamma-ray flux measurements of nearby supernovae and neutron
stars.  Upcoming measurements planned with the GLAST satellite may
improve bounds further or lead to a new discovery. Both sets of
constraints are weakened if the dimensions are of unequal size, if
there are additional fast decay channels such as other branes for
the KK gravitons to decay into, or if graviton emission is
suppressed as in reference \cite{dgkn}.

From the four-dimensional point of view, the higher dimensional
graviton with momentum in the extra dimensions appears as a
massive particle, leading to a sum of Yukawa potentials from the
tower of Kaluza-Klein (KK) modes in addition to the massless
graviton potential.  At distances or order $R_n$, only the lowest
massive mode contributes significantly while the higher modes are
exponentially suppressed.  For distances $r<<R_n$, many modes
contribute and change the power law dependence of the force from
the Newtonian $1/r^2$ to $1/r^{2+n}$.  Corrections to the
newtonian potential between two masses $m_1$ and $m_2$ are
typically parameterized according to the form
\[V(r) = -G_N
\frac{m_1 m_2}{r} (1 + \alpha e^{-r/\lambda})
\]
where $\alpha$ and $\lambda$ characterize the strength relative to
gravity and range of the new force, respectively.  For distances
of order $R_n$ or greater, the range $\lambda$ is the inverse of
the lightest KK mass and the strength $\alpha$ equals its
degeneracy \cite{kands,fandl}.  At shorter distances, more massive
KK modes contribute until eventually the power law behavior of the
force changes.  We adopt the convention of references
\cite{kands,fandl} and consider only the leading term
corresponding to the lightest massive KK modes. For example in the
cases of toriodal and spherical compactification \cite{kands}
\[
V(r)_{\rm{n-torus}} = -\frac{G_N m_1 m_2}{r} (1+2n_0 e^{(-r/R_0)})
\]
\[
V(r)_{\rm{n-sphere}}= -\frac{G_N m_1 m_2}{r} (1+ (n+1)
e^{(-\sqrt{n}r/R)})
\]
for $n_0$ equal radii of size $R_0$.

Due to the stringent astrophysical constraints on two- or three-
equal sized large extra dimensions, and their even smaller size
for $n>3$, it is unlikely in this case that the KK gravitons can
be observed at table-top experiments.  However, we stress that
these constraints strictly apply to the case of equal extra
dimensions. If the extra dimensions are not of equal size, it is
possible some of the dimensions may be large enough to be
detected.  To illustrate this we plot the cases in
$(\alpha,\lambda)$ space for toroidal compactification with one or
two large radii (amongst possibly many smaller radii).

\subsection{Gauged Baryon Number in the Bulk}

If in addition to the graviton there are bulk gauge particles,
their effective four-dimensional gauge coupling $g^2_4$ can be
many orders of magnitude stronger than gravity \cite{add3}. As a
particular case we consider gauged baryon number $B$, with the
gauge symmetry spontaneously broken only on a different
submanifold than our own.  As discussed in references
\cite{add3,nimasavas}, such a situation can lead to an enormous
suppression of the proton decay rate. In the following we
systematically explore the parameter space for the force strength
and range.  Several details are deferred to the Appendix.

Since ordinary matter is primarily composed of baryons, the mass
of a macroscopic object is roughly in proportion to the number of
baryons, apart from small effects due to binding energy and the
electron mass. The expected ratio of the gauge to gravitational
forces will take the form \be \alpha_g =
\frac{F_{gauge}}{F_{grav}} = \frac{g^2_4}{4\pi}\frac{1}{G_N
m^2_p}\label{alpheq1} \ee where $m_p$ is the mass of the proton.
The effective four-dimensional gauge coupling is related to the
massive $4+n$-dimensional coupling by the volume of the extra
dimensions
\[\frac{1}{g_4^2} = \frac{V_d}{g^2_{(4+n)}}.\]  We can express the
(4+n) dimensional coupling in terms of an ultraviolet cut-off
scale for the gauge theory $\Lambda$ which we expect to be of
order $M_{*}$: $g^2_{4+n} = \Lambda^{-n} \O_{4+n} \rho$, where
$\O_d = \Omega_{(d-1)}/{(2\pi)^d} $ is the 1-loop suppression
factor. If the coefficient $\rho$ is ${\mathcal{O}}(1)$, this
signifies strong coupling in the 4+n dimensional theory, as loop
effects become comparable to tree-level.  Expressing the
4-dimensional coupling in eq. (\ref{alpheq1}) in terms of $\Lambda
\sim M_{*}$, we find the baryon-number force can easily reach
strengths of $10^6-10^8$ times gravity, and even higher magnitude
for strong-coupling with a large number of extra dimensions.

To avoid conflict with experiment, the baryon number gauge field
must acquire a mass.  If the gauge symmetry is spontaneously
broken by a scalar field $\chi$ obtaining a vacuum expectation
value $\langle \chi \rangle $, the resulting mass of a gauge
particle $A_\mu$ becomes $m_A = g_4 \langle \chi \rangle $.  Here
again we assume that $\chi$ condenses on a brane other than our
own.  It was shown in reference \cite{dvali1} that forces mediated
by bulk gauge fields can be exponentially weaker than gravity if
the bulk gauge symmetry is spontaneously broken on our brane.  For
$\langle \chi \rangle = \beta~M^{*}$ with $\beta$ of
${\mathcal{O}}(1)$, the Compton wavelengths are in an interesting
range for sub-millimeter experiments.  A range of predicted
parameter space is provided along with a more detailed analysis in
the Appendix. A portion of the allowed phase space appears in
Figures 1 and 2 for comparison with other sub-millimeter forces
and experimental bounds.

Astrophysical bounds similar to those discussed for gravitons
\cite{add3},\cite{raffelt},\cite{cullenmaxim} can also apply for
gauge particles in the bulk.   In Supernova 1987A, about $10^{53}$
ergs of gravitational binding energy was released in a few
seconds. One requirement is therefore that the total luminosity of
Kaluza-Klein particles does not exceed $\sim 10^{53}~
\rm{erg~s}^{-1}$.  The temperature of the supernova is
approximately $T\sim 30 ~\rm{MeV}$, and most KK particles are
produced with an energy of order $\sim 100~ \rm{MeV}$.  The
constraints on the total luminosity for KK gravitons imply $M_{*}
> 30$ TeV for $n=2$ extra dimensions \cite{add3}.  For the case of
gauge particles in the bulk, we expect the same amount of energy
that would have gone into KK gravitons to now produce KK gauge
bosons, producing roughly the same number of particles.  However
the rate of production for gravitons goes like $T^n/M_{*}^{2+n}$
and for gauge bosons like $T^{n-2}/M_{*}^n$, which is more rapid.
Therefore the constraints on the fundamental scale due to graviton
emission for $n$ extra dimensions apply for gauge particle
emission with $n+2$ dimensions.  A similar situation is discussed
in reference \cite{nimasavas} for the case of bulk scalars. This
implies the more stringent limit of $M_* > \sim 30$ TeV for the
case of $n=4$ extra dimensions.

 {\bf{Neutron Star Limits.}} Kaluza-Klein excitations of the
gauge bosons can also become gravitationally trapped and remain
for some time in a cloud surrounding neutron stars.   Their
subsequent decays into photons can in certain cases produce
observable gamma-ray signals detectable by EGRET.  The decay width
for such particles can be roughly computed as \be \Gamma =
\frac{M_*^2~T}{M_{\rm{Pl}}^2}~, \label{lifetime} \ee where $T$ is
the temperature, typically of order $30 ~\rm{MeV}$ for a
supernova. The lifetime becomes $\sim 10-10000$ years for $T$ from
$1$ GeV-$1$ MeV.  (The decay width for gravitons goes as $\Gamma
\sim T^3/M_{\rm{Pl}}^2$, leading to lifetimes of order $6 \times
10^9$ years.) However, the situation can be quite different
depending on the symmetry that is gauged. For $B-\zeta L$, where
nonzero $\zeta$ denotes an admixture of lepton number, there are
decay channels into neutrinos.  In this case eq. (\ref{lifetime})
is a good approximation and the decays can occur on a time-scale
of $10^2$ years.  Such a situation provides little direct
observable gamma signal for EGRET.  Also KK annihilation into
positrons is possible, but the resulting gamma rays from
positronium annihilation are of too low energy to be a useful
EGRET source.  In the case of pure $B$, where $\zeta=0$, the
lifetime given in eq. (\ref{lifetime}) has to be amended since the
decay into photons occurs only at higher order.  The decay width
in this case is multiplied by an additional factor of
$(\alpha/2\pi)^2$ due to a virtual fermion loop. This increases
the lifetime by a factor of $\sim  10^6$, making it more
comparable to the graviton case at $\sim 1.3 \times 10^8$ years.
The resulting improvement in the bound on $f_{KK}$ as compared
with the graviton case is about 50 times.  For bulk gauge
particles, the bound on the compactification scale varies with
$f_{KK}$ as $M_{*{\rm{min}}}~\alpha~(f_{KK})^{-1/n}$.  In this
situation, the bounds that applied for $n$ extra dimensions in the
graviton case now apply to $n+2$ extra dimensions.  However, even
in the case of pure B further limits cannot be derived from these
neutron star gamma rays and excess heat due to direct
re-absorption of the KK gauge particles through inverse
bremstrahlung, which occurs on a rapid time scale and is not
loop-suppressed.  In this way, there would not be enough remaining
KK gauge particles to contribute to heating or gamma ray limits.
KK re-absorption for the case of gravitons was taken into account
in reference \cite{raffelt2} and in this case the results did not
appreciably change the graviton limits quoted in reference
\cite{raffelt}.  As was the case for graviton decays, the
constraints may be even further weakened if the dimensions are of
unequal size, or if there are additional fast decay channels such
as "photons" on other branes for the KK particles to decay into.

\subsection{Yukawa Messengers}
An additional possibility is that scalar particles may inhabit the
bulk of extra dimensions and mediate macroscopic forces, such as
the Yukawa messengers considered in reference \cite{nimasavas}.
The messenger fields could be responsible for communicating flavor
symmetry breaking from other branes to our brane, and for example
can attribute the weakness of light generation Yukawa couplings to
geometrical power-law suppression or exponential suppression due
to the mass of the messenger fields.  In this framework, the vast
variation in strengths of the Yukawa couplings is recast as a
variation in distances to other branes.  However, even if the
messenger fields do not condense on our brane, they can still
mediate forces much stronger than gravity.  If the messenger
fields acquire mass due to supersymmetry breaking on our wall,
their Compton wavelength can be in the sub-millimeter range.  As
was the case with gauged baryon number, the coupling strength for
such forces due to the zero mode can be large even if the extra
dimensions are small enough to make the Kaluza-Klein modes too
heavy to be detected in sub-millimeter experiments.  The coupling
strength $\rho \sim v/M_{\rm{Pl}}$ where here $v$ is the Higgs
vacuum expectation value of $174~\rm{GeV}$ and here $M_{\rm{Pl}}$
is the reduced Planck scale of $2.43 \times 10^{18}~\rm{GeV}$.
Comparing to gravity, $\rho^2/(G_N m_{\rm{nucleon}}^2) \sim 10^6$.
As with the vectors described in the previous section, similar
astrophysical constraints apply to the scalars in the bulk as
discussed in reference \cite{nimasavas}.

\section{Using Bose-Einstein condensed atoms to probe (sub)-micron distances}

In recent years the field of atomic interferometry has produced a
series of amazing measurements, including extremely high precision
measurements of the acceleration due to the Earth's gravity at the
level of a part per billion or better.  These advances have been
made possible due to the remarkable techniques developed for
trapping and cooling alkali atoms (see, for example, the 1998
Nobel lectures of Chu, Cohen-Tannoudji, and Phillips
\cite{nobel1}). Among recent experiments have been a series of
atomic-fountain type measurements where an atomic beam is launched
upwards and allowed to accumulate a phase shift in the Earth's
gravitational field in a Mach-Zehnder-type interferometer
configuration \cite{pcc}. Similar experiments have also been
carried out to perform sensitive measurements of gravity gradients
\cite{mffsk}. Such precision techniques can also be used to
measure large deviations from Newton's constant at short
distances.  Among the challenges of applying such systems to study
gravity from nearby macroscopic objects is obtaining the optimized
beam size and divergence necessary to allow a short range
interaction to be carried out in a systematic way over a long
enough time period. Typical spatial extents and velocity spreads
are of order mm and cm/s respectively, making sub-micron
experiments difficult. Experiments involving atoms trapped at
fixed separation from a source mass surface are in this respect
preferable.

Since its first experimental realization in 1995, Bose-Einstein
condensation of alkali gases has become a widely growing area of
research (see, for example, the 2001 Nobel lectures of Cornell,
Ketterle, and Wieman \cite{nobel2} or reference\cite{leggett} for
a recent review). Interference of atomic de Broglie waves which
develop a relative phase shift due to the Earth's gravity has been
observed in vertical arrays of trapped Bose-Einstein condensed
atoms \cite{andersonkasevich98}. The traps were located at the
antinodes of a laser standing wave, with the trap well depth
determined by the laser intensity.

In the following we describe a setup involving arrays of
Bose-Einstein condensed atoms trapped nearby a surface at the
nodes or antinodes of a laser standing wave.  The traps are thus
loaded with atoms in coherent superpositions of states localized
at differing distances from the surface.  For each potential well,
the de Broglie phase of the center of mass wave function of the
atoms evolves according to the interaction potential of the local
environment. The wave function of atoms localized at different
potential wells of the laser can accumulate a differential phase
shift due to the distance-dependence of the interaction potential
with the wall. The potential will generally be a superposition of
the Casimir-Van der Waals potential along with other backgrounds
and possibly new short-range interactions. By adequately
subtracting out the Casimir interaction and other background
interactions as we discuss below, significant improvements can be
made over previous searches for new forces below 1 micron.  In
particular the improvements could be 6 to 9 orders of magnitude at
1 micron, allowing forces of $1000$ to $10^6$ times gravity to be
detected at these distances.  Such short length scales have been
relatively inaccessible to tabletop torsion balance and
micro-cantilever experiments due to the necessity of having nearby
moving macroscopic mechanical parts and the unfavorable scaling of
the gravitational force with their size.

\subsection{Experimental Setup and Geometry}
\begin{figure}
\includegraphics[width=1.0 \columnwidth]{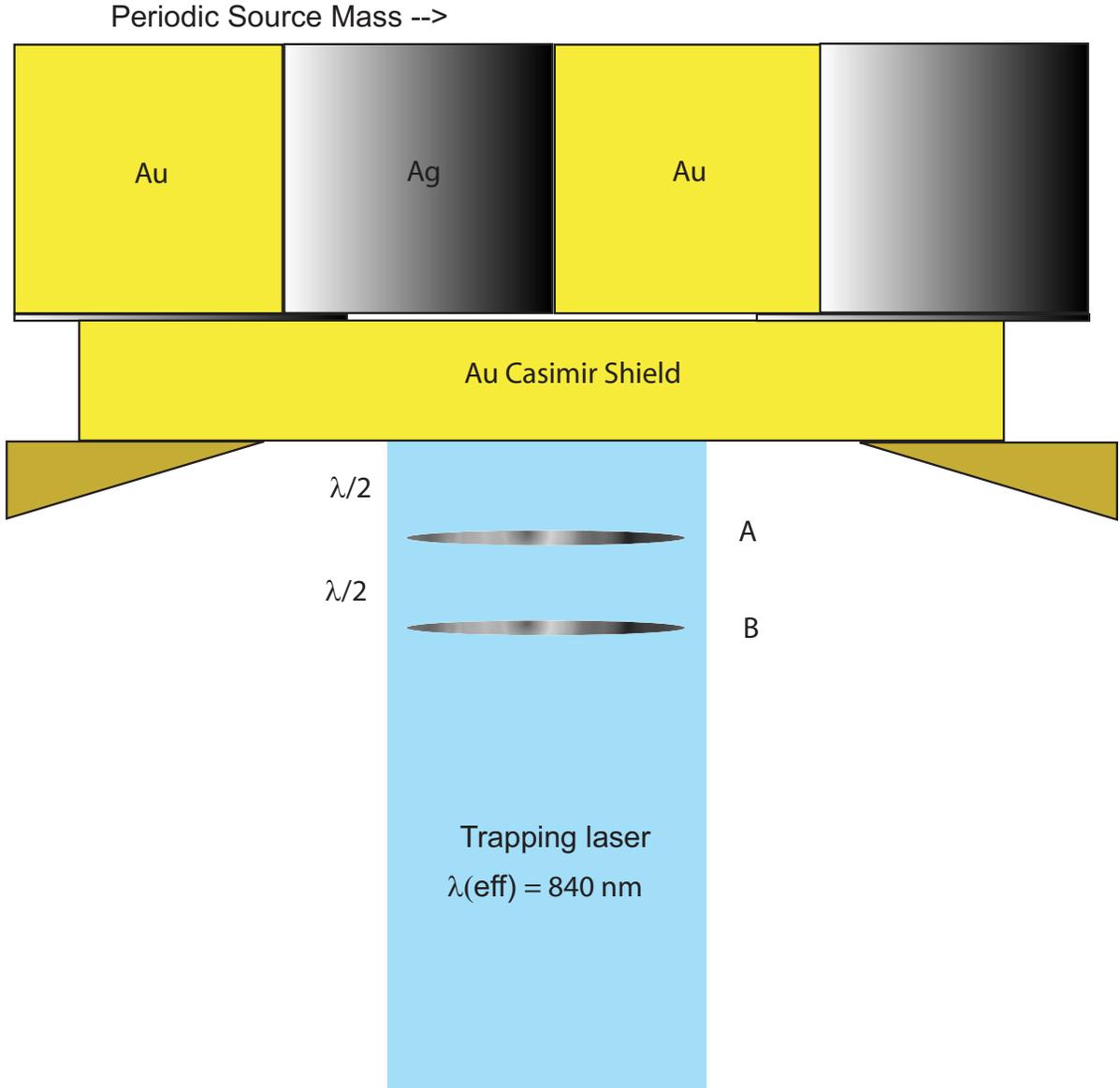}
\caption{Proposed experimental arrangement.  The standing wave
cavity formed by reflection from the Au Casimir shield is used to
trap BEC atoms in the two potential wells nearest to the periodic
source mass array.} \label{markfig}
\end{figure}

We consider $^{87}$Rubidium atoms prepared in coherent
superpositions of states localized at the nodes or antinodes of a
standing wave of an infrared laser of wavelength
$\lambda_{\rm{las}}$.  For a laser wavelength that is red de-tuned
from the dominant Rubidium D2 line, the atoms are attracted to
regions of high laser intensity, corresponding to trapping at the
anti-nodes. On the other hand, for blue de-tuned light the
potential minima occur at the nodes of the standing wave.  In the
first case the potential wells will be centered at distances
($\lambda_{\rm{las}}/4$, $3\lambda_{\rm{las}}/4$,...) from the
surface, in the case of normal laser incidence.  For blue
de-tuning the wells are located at distances
($\lambda_{\rm{las}}/2$,$\lambda_{\rm{las}}$,...) from the
surface.  The effective well separation and surface separation can
readily be made larger in the case of oblique incidence at angle
$\theta$, where $k_{\perp}^2 < k^2$ and the effective wavelength
determining the trap spacing becomes $\lambda =
\lambda_{\rm{las}}/\cos{\theta}$. Thus a variety of surface and
well separations are attainable depending on the trap geometry.
The atomic well depths can be adjusted by varying the laser
intensity to overcome the atom-surface interaction potential and
the Earth's gravitational field. Transverse confinement can be
achieved for example through the gaussian envelope of the beam in
the case of red de-tuning, or with additional laser beams.

For definiteness we consider a standing wave with an effective
trap spacing of $\lambda/2=420~\rm{nm}$. We also take the
separation of the first trap and the surface to be $420$ nm. The
proposed geometry is illustrated in Figure \ref{markfig}.  The
laser is reflected from a 420 nm thick shield of gold.  The
periodic source mass consists of alternating regions of more and
less dense material, for example gold and silver. Although less
dense materials are preferable for the Yukawa force contrast,
silver is chosen for the similarity of its diamagnetic response to
that of gold to ameliorate possible problems with magnetic
backgrounds. The source masses are taken to be 100 microns wide by
100 microns deep by 10 microns tall. The source mass can be moved
as a whole laterally by a piezo-electric device over several
hundreds of microns.  To reduce temperature changes in the Casimir
shield as the source mass moves, a very small space is left open
directly behind the shield.  We consider a population, by $\sim
10^6$ atoms, of the first two potential wells closest to the
surface, denoted by A and B, respectively.  The atoms will be
spread out in a pancake-like configuration with a transverse
extent of a few microns.  The wells must be loaded with a fixed
initial relative phase between the two parts of the condensate.
The optical wells can be loaded for example by ramping up the
laser field after evaporative cooling. The population of only the
first two wells can be achieved perhaps through magnetic or
optical techniques to translate the lattice closer to the shield
wall until the destruction of any extra occupied wells occurs. The
de Broglie phase of the wave function localized at the first well
(A) will evolve more rapidly than that of the wave function at the
second (B) due to the atom-surface interaction potential. After an
interrogation time of 1-10s, the laser intensity can be rapidly
turned down, allowing the wave-packets to escape, spread out, and
overlap spatially. The resulting interference pattern can then be
detected using optical fluorescence. If the accumulated phase
difference is due to a Yukawa-type interaction with the surface,
the phase difference will change depending on whether a silver or
gold section of the sense mass is positioned behind the Casimir
shield. A series of such experiments can be performed such that
between each experiment the source mass pattern is moved laterally
behind the screen. Depending on whether a more or less dense
region of material is behind the shield, the resulting phase shift
will display a periodic behavior. On the other hand, the
Casimir-Polder interaction will be largely the same due to the
gold Casimir shield. Also, any patch fields that contribute to the
interaction can be rejected as a common mode.  The magnitude of
the Casimir-Polder interaction at 420 nm is some $8$ times the
gravitational interaction with the Earth. Due to its gigantic
magnitude, it is crucial then to consider the finite-thickness
corrections to the Casimir potential to estimate the potential
reach of the experimental setup.

\subsection{Sensitivity}
{\bf{Scaling.}}  We compute the difference in the frequency at
which the de Broglie phase evolves for adjacent populated wells of
the laser near the surface due to a Yukawa-type potential of
strength $\alpha$ and range $\lambda$. Taking the thickness of the
source masses to be $r$ and $2r$ and the atom-wall separation to
be $r$, the potential difference can be roughly approximated as
\be \delta V(r) \sim 2\alpha G_N m \rho r \delta{r} \ee where we
consider a Yukawa potential of range $\lambda \sim r$ and take
$\delta r \sim r$, so that we have \be \delta V(\lambda) \sim
2\alpha G_N m \rho  \lambda^2. \label{sens} \ee Here $m$ denotes
the mass of an individual atom.  For example, taking Rubidium-87
and a gold wall, the corresponding frequency shift for $\lambda=1
\mu$m is $1.5 \alpha \times 10^{-10}$ Hz.  To obtain a more
precise estimate, we numerically integrate the Newtonian plus
Yukawa potential.

With $\sim 10^6$ atoms in the condensate, we estimate the minimal
detectable phase shift per shot as $10^{-3}$ radians.  For a
conservative estimate of $1$ s interrogation time, the minimal
resolvable frequency shift is $1.6 \times 10^{-4}$ Hz. This
corresponds to an acceleration sensitivity of roughly $10^{-7}$ g,
where $g$ is the acceleration of the Earth's gravitational field
at the surface, where we again take a trap spacing of
$\lambda/2=420~\rm{nm}$. Ultimately, an improvement could be
obtained by an averaging over $10^4$ shots, allowing the phase
shift to be detected at the $10^{-5}$ level. This along with
increasing the interrogation time to $10$ s yields a minimal
detectable frequency shift of $1.6 \times 10^{-7}$ Hz,
corresponding to an acceleration sensitivity of order $10^{-10}$
g.  Obtaining larger interrogation times may be difficult due to
loss of coherence from collisions or laser instability
\cite{mark}. In the following we assume a sensitivity of $1.6
\times 10^{-4}$ Hz. Also, the systematic effects we consider in
the following sections are not generally problematic at this
level, however may be more challenging for a measurement with
$10^{-7}$ Hz sensitivity as will be discussed.

Using this minimal detectable frequency shift and the numerical
results for the Yukawa potential, we generate a plot of the
alpha-lambda reach in Figures 1 and 2.  Such an experiment is
particularly favorable in the sub-micron length scales, where
macroscopic cantilever and torsion balance experiments become
increasingly more challenging.

For the geometry described above with a BEC-surface separation of
0.42 microns and well separation of 0.42 microns, we evaluate the
frequency shift of the potential wells due to a Yukawa potential
of strength $\alpha$.  For lambda of $1 \mu$m, we find $1.2 \times
10^{-10} \alpha$ Hz for the frequency shift, allowing alpha of
$\sim 10^6$ to be probed. A plot of the projected sensitivity on
alpha-lambda space appears in Figures 1 and 2 as a solid green
line.  We also indicate the projected improvement possible with
the sensitivity taken to the $10^{-7}$ Hz level as a solid red
line.  In Figures 1 and 2, the green curve is calculated for a
surface separation, trap spacing, and shield thickness of 420 nm
with 1 second interrogation time and a single shot.  In Figure 2
the red curve is also computed at a well separation, surface
separation, and shield thickness of 420 nm. The red line in Figure
1 has been computed for a shield thickness, trap spacing, and
surface separation scaled from 420 nm to 600 nm. Here we assume 10
seconds of interrogation time and average over $10^4$ shots,
corresponding to $10^{-7}$ Hz sensitivity. The larger scale of 600
nm becomes advantageous for measurements near $\lambda = 1\mu$m as
it can better suppress the Casimir background.  We note that for
these geometries the sensitivity levels off above 1 times gravity
at large lambda. By scaling the geometry and trap spacing
together, the estimated sensitivity roughly follows
Eq.(\ref{sens}) so that $1 \times$ gravity is achievable at $\sim$
30 microns. These larger length scales are also accessible by
upcoming micro-cantilever or torsion oscillator experiments.
Therefore in figures \ref{submm} and \ref{subum} we have
emphasized the reach at the micron scale and below.

\subsection{Systematics}
\subsubsection{Atomic interactions}
Atomic interactions in general will also produce differences in
the chemical potential of the two parts of the condensate, leading
to relative phase differences that would occur in addition to
those caused by the differential gravitational potential.  As one
possible solution, care could be taken to have an approximately
equal number of atoms in the two clouds, perhaps by using 2-d
lattice configurations, where arrays of single atoms are confined
transversely as well as longitudinally \cite{mark}. Alternatively
the Feshbach resonance could in principle be used to highly
suppress atomic interations \cite{feshbach}. For example, it has
been shown that the s-wave scattering length can in this way be
tuned over several orders of magnitude and set effectively to
zero, thus turning off the chemical potential due to atomic
interactions.
\subsubsection{Thermal fluctuations}
Finite temperature fluctuations of the BEC phase may present an
additional experimental challenge, common to interferometry using
BEC.  For example, such effects have been studied theoretically
and experimentally \cite{thermal} for highly-elongated condensates
where they are shown to become problematic, as the system becomes
quasi-one-dimensional.  Care should be taken to keep a three
dimensional nature to the condensate to minimize this effect. Also
a finite temperature in the surrounding {\it{materials}} may pose
an experimental challenge.  Thermal currents in metals in the
local environment can produce magnetic field fluctuations which
can limit the condensate lifetime \cite{surface}.  Such
fluctuations can be minimized by lowering the system temperature,
using lower-conductivity materials, and minimizing the thickness
and transverse area of the reflecting shield.
\subsubsection{Casimir Background} {\bf{Finite Thickness and
Conductivity.}} The frequency shift due to the bare Casimir force
is quite large (on the order of $8$ kHz). However, in the case of
infinite conductivity, the {\it{differential}} frequency shift
between regions of metal of varying thickness is zero.  In
practice, the finite conductivity of the metal as well as its
finite thickness has to be taken into account. It was shown in
reference \cite{price} for the case of two metal walls that the
differential Casimir background due to differences in thickness
rapidly dominates the Yukawa force as thickness and plate
separation decreases, thus making measurements of gravitational
strength Yukawa forces difficult below a few microns.  The
geometry studied consisted of a semi-infinite probe mass of finite
conductivity separated by a distance $D$ from a finite
conductivity source mass of thickness $D$.  The result for the
Casimir force and a gravitational strength Yukawa force of range
$D$ was then compared to the case source mass thickness $2D$. It
was shown that the differential Casimir background due to
differences in thickness becomes comparable to the differential
Yukawa force at distances of about $3$ microns and rapidly
dominates the Yukawa force below this length scale.  In this work
we show that by replacing the metal probe mass with a dielectric,
or with an atom in particular, the differential Casimir force is
considerably smaller, making the domination over the Yukawa
potential less severe at sub-micron lengths. Following references
\cite{price}, \cite{lambrecht} we employ a reflection-based model
for computing the Casimir force between two walls.  We obtain
similar results to reference \cite{price} for metallic walls.
Also, we obtain a reduction factor $\eta_F$, (defined precisely in
reference \cite{lambrecht}) which describes the reduction of the
Casimir force at small separations and for finite conductivity,
that agrees well with the results of that reference in the case of
semi-infinite walls.  The parameter $\eta_F$ is defined as \be F_C
= \eta_F F_P \ee where $F_{P}$ is the perfect conductor result.
The integral expression \be \eta_F = \frac{120}{\pi^4}
\int_0^{\infty} dK K^2 \int_{0}^{K} d\Omega \sum_p
\frac{r_p^2}{e^{2K}-r_p^2} \ee gives $\eta_F$ in terms of the
reflection amplitudes $r_p$ which can depend on the polarization,
conductivity, frequency, and wall thickness \cite{lambrecht}. Here
$K=\kappa L $ and $\Omega = \omega \frac{L}{c}$ are the wavenumber
and frequency measured with respect to the cavity length $L$.

The Casimir force between the mirrors can be expressed in terms of
the imaginary part of the dielectric function of the walls. For
metallic walls, the dielectric function is
\[
\epsilon(i\omega) = 1 + \frac{(\omega_p)^2}{\omega(\omega+\gamma)}
\]
where we assume a Drude model for the metals, and $\omega_p$ and
$\gamma$ are the plasma frequency and relaxation frequency,
respectively.  The magnitude of the Casimir force changes rapidly
as the length scales of thickness and separation approach the
plasma wavelength $\lambda_p$.  The expression for the reduction
factor $\eta_F$ which describes the fraction of the perfect metal
Casimir result can be written as an integral over all frequencies
and wavenumbers. In particular, the low-frequency response of the
dielectric function of metals diverges as $\omega \rightarrow 0$.
By replacing one of walls with a dielectric $\epsilon$, the
low-frequency response becomes weaker and contributes less to the
Casimir force.  As a consistency check, we evaluate the expression
for a dielectric wall and metal wall using both the reflection
model of reference \cite{lambrecht} and by numerically integrating
the zero-temperature Lifshitz result directly\cite{lifshitz}.  The
two numerical calculations agree to a part in $10^5$.  After
demonstrating the equivalence of the two models for the
semi-infinite case of dielectric and metal walls, we proceed with
the reflection model to study the finite thickness dependence. We
find the result is less sensitive to the thickness of the source
metal, which improves the situation considerably for new force
detection below 3 $\mu$m. The reduction factor for the Casimir
Force $\eta_F$ for the two walls is computed for the case of wall
thickness $D$ and $2D$, for wall separation $L$.  We define the
quantity \be \Delta \eta = \frac{\eta_F^{(2D,L=D)} -
\eta_F^{(D,L=D)}}{\eta_F^{(D,L=D)}} \ee which expresses the
fractional differential Casimir force for the source walls of
different thickness.  We list a table of values of $\Delta \eta$
for probe walls of metal and dielectric materials below.  We also
note that the interaction between two dielectric walls is much
more sensitive to their thickness, due to the lack of screening
present in metals.  For example, we find $\Delta \eta$ can be as
large as 1 percent for $D=1~\mu\rm{m}$ and $\epsilon = 10$, which
compares rather poorly with the metal-dielectric case.

\begin{table}
\caption[]{Estimates for the differential Casimir force between a
gold source mass of thickness D versus 2D, and a semi-infinite
probe mass of varying materials.  Smaller values indicate less
sensitivity to the differential thickness.} \vglue0.3cm
\begin{center}
\begin{tabular}{|c|c|c|c|}
\hline
$D$ & $\Delta \eta_{\rm{metal-metal}}$ & $\Delta \eta_{\rm{metal-dielectric}}$ & $\Delta \eta_{\rm{metal-dielectric}}$  \\
$(\mu m)$ & $$ & $\epsilon = 100$ & $\epsilon = 1.001$ \\
\hline \hline
 & & & \\
$.1$ & $9 \times 10^{-5}$ & $3 \times 10^{-5}$ & $2 \times 10^{-5}$  \\
$.3$ & $5 \times 10^{-6}$ & $1 \times 10^{-9}$ & $9 \times 10^{-11}$  \\
$.6$ & $2 \times 10^{-6}$ & $1 \times 10^{-10}$ & $4 \times 10^{-12}$  \\
$1$ & $8 \times 10^{-7}$ & $1 \times 10^{-10}$ & $4 \times 10^{-13}$  \\
 & & & \\
\hline
\end{tabular}
\end{center}
\end{table}

To achieve the limit of the atom-wall interaction, we consider
rarifying the dielectric medium.  This technique was used by
Lifshitz to derive the individual atomic Van der Waals potential.
We also add the dominant resonance for Rubidium D2 line. The
dielectric function satisfies
\[
\epsilon(i\omega) = 1 + 4\pi n { }\alpha(i\omega)
\]
where $\alpha$ is the dynamical polarizability and $n$ is the
number-density of atoms. Considering the dominant D2 line at 780
nm with an oscillator strength of nearly 1, we approximate the
polarizability as
\[
\alpha(i\omega) = \frac{\omega_0^2{ }\alpha_0}{\omega_0^2 +
\omega^2}.
\]
The role of the parameter $\epsilon(0) - 1$ is now played by the
quantity $4 \pi n \alpha_0$.  For $^{87}$Rb, we have $\alpha_0 =
2.7 \times 10^{-23}$cm$^3$, and even for a high number density of
order $n \sim 4 \times 10^{16}~$cm$^{-3}$ we obtain $\epsilon(0)
-1$ of $\sim 10^{-5}$, indicating even more favorable scaling of
$\Delta \eta$ than shown in the Table for $10^{-3}$.

\begin{figure}
\includegraphics[width=1.0 \columnwidth]{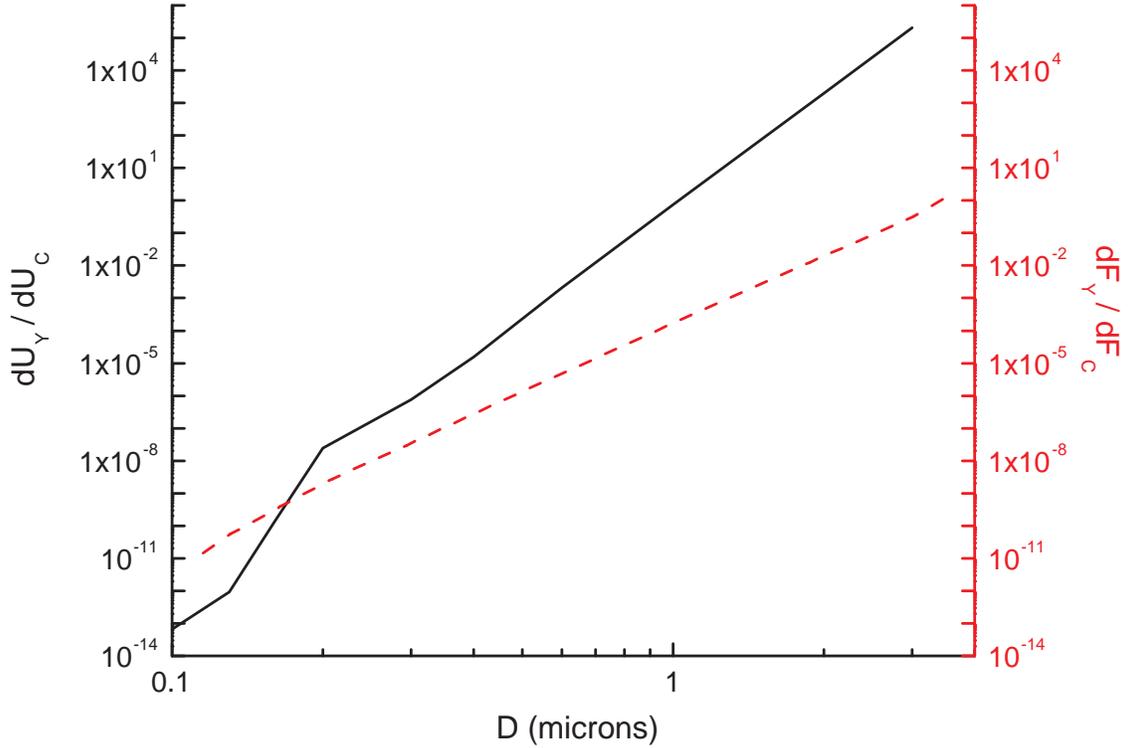}
\caption[]{Estimate for the Yukawa detection limit due to
Casimir-Van der Waals background for Yukawa range $\lambda = D$
and strength relative to gravity $\alpha=1$ is shown as the black
solid line for the atom-wall system. The well spacing and
atom-wall separation are also $D$, and we compare regions of metal
thickness $D$ and $2D$. The metal-metal case studied in reference
\cite{price} is shown as a red dotted line for comparison.}
\label{fig3}
\end{figure}

It remains to estimate the minimal detectable alpha due to a
Yukawa potential limited by the differential Casimir force.  The
Casimir potential due to the atom at distance $r$ from an
infinitely conducting surface can be written
\begin{equation}
U_C = -\frac{3 \hbar c} {8 \pi} \frac{ \alpha_0}{r^4}.
\label{at-m}
\end{equation}
In practice we expect corrections due to the finite conductivity
and the dynamical polarizability of the atom.  The finite
conductivity correction is less than a factor of 2 reduction for
the length scales of interest.  The equation (\ref{at-m}) is
strictly valid in the limit of large separation, at length scales
greater than $\lambda /2\pi$ where now $\lambda$ is the wavelength
which contributes to the atom's polarizability.  For lengths below
this scale, the Casimir screening due to retardation becomes less
effective and the power law changes to $1/r^3$ corresponding to
the van der Waals interaction.  For Rb, the dominant wavelength is
$780$ nm and since we are interested in length scales above $100$
nm, the form in Eqn. (\ref{at-m}) is a reasonable approximation.
We now multiply the parameter $\Delta \eta$ by the difference in
Casimir potentials of the wells obtained through Eq. (\ref{at-m})
to obtain an estimate of the difference in potential shift due to
the metal walls of the two thicknesses.  The resulting $\Delta
\delta U_C$ we compare with $\Delta \delta U_{\rm{Yukawa}}$. (Here
$\Delta$ signifies the change from a region of thickness $D$ to
$2D$, and $\delta$ signifies the different locations of the
condensates A and B. In Figure \ref{fig3} we abbreviate both by $d
= \Delta \delta$.) In order to illustrate the scaling, we consider
a set-up with equal length $D$ for $\lambda$, the well spacing,
and the atom-wall separation, and compare regions of metal
thickness $D$ and $2D$. We display the results in Figure
\ref{fig3} and take a Yukawa force of gravitational stength,
$\alpha=1$. For comparison we include as a dotted line the
estimate adapted from reference \cite{price} which illustrates the
scaling of the differential $\alpha=1$ Yukawa and Casimir forces
for a metallic probe wall in place of the atoms. The situation for
the atom-wall setup is more optimistic than that of the metallic
wall-wall setup (shown in Fig [5] of reference \cite{price}) by
several orders of magnitude between $200~\rm{nm}$ and 3 microns.
In Figure \ref{fig3} there are two separately labelled vertical
axes shown since the atom-wall Yukawa detection limit due to the
differential Casimir background is determined by the ratio of the
potential differences, whereas for the metallic wall-wall system
the relevant quantity is the ratio of the forces. This does not
prevent a direct comparison of the $\alpha$- reach of the two
systems however.  For example, we see from the figure that
$\alpha$ of $10^3$ can be reached at .6 microns in the metal-atom
case and at about 1.5 microns in the metal-metal case.

{\bf{Temperature Effect.}} The temperature dependence of the
Casimir force becomes quite weak for low temperatures and small
separations.  However, it is important to estimate the
temperature-dependent contribution since the full magnitude of the
Casimir force is so large.  A perturbation approach has been
developed in reference \cite{panenko}. We employ the parallel
metal plate expression valid for separations below 2 microns as a
function of temperature T:
\begin{equation}
\Delta \eta_F (T) = (\frac{4}{3t^5} + \frac{2\delta}{a}
\frac{15}{\pi^2} \frac{3 \zeta(3)}{t^4})\Delta t...,
\end{equation}
where $t = T_{\rm{eff}}/T$, $T_{\rm{eff}} = \hbar c / 2 a k_B$,
$\delta = \lambda_p /2\pi$, and $a$ is the plate separation.
Taking a separation of $a=840$ nm, a gold surface of $\lambda_p =
136$ nm, we find for a room-temperature of $295$ K, the parameter
$t=4.6$. This amounts to a fractional change in the Casimir force
$\Delta \eta_F = .0012 \Delta t$ and writing $\Delta t$ as
$t\frac{\delta T}{T}$, we have
\begin{equation}
\Delta \eta_F (295{\rm{K}}) = 0.0055 \times \frac{\delta T}{T}...
\end{equation}
which can be a significant effect.  If the entire reflecting
surface changes temperature by $10^{-6}$ K during the
measurements, the frequency shift is at the $10^{-7}$ level.  For
the proposed initial parameters the sensitivity to wall
temperature is at the $\sim$ mK level. However, most of the
Casimir force results from the wall material closest to the atom,
which we suggest to take as a shield of uniform material, and have
the pattern of varying density source masses separated from this
material by a small distance to prevent heating upon motion.  Even
though a small temperature gradient could be supported across the
shield wall, it is unlikely that the shield surface temperature
will vary periodically with the source mass density, as the two
are not in direct thermal contact. A periodic temperature gradient
in the source mass distribution itself changes the expected
Casimir force very little, as its effect comes in only at the
level of $\Delta \eta$ due to the thickness of material calculated
in the previous section. As an additional handle, the temperature
of the surfaces can be controlled externally to quantify the
effect experimentally. Also, decreasing the temperature improves
the situation considerably, yielding a sensitivity to a full wall
temperature change of $ \sim 10^{-3}$ K at $77$ K and only to $
\sim 10$ K at liquid helium temperature for frequency shifts of
$10^{-7}$ Hz.

 {\bf{Isotope Effect.}}  A unique feature of atomic systems
in contrast to macroscopic objects is that the electrical
properties and mass of the atom can be toggled in a precise way by
taking advantage of other stable atomic isotopes.  For example, by
using $^{85}$Rb, and comparing to an experiment done with $^{87}$
Rb, one expects the force to change at the $10^{-2}$ level, while
the Casimir force changes only at the $10^{-4}-10^{-5} $ level. To
verify this claim, we compute the Casimir force for each isotope
according to the Lifshitz model, and simply change the wavelength
of the dominant transition from $780$ nm to $780.1$ nm. Although
this technique decreases the sensitivity to alpha by $2$ orders of
magnitude, it may be useful for doing measurements at around $100$
nm separation from the surface, where the differential Casimir
force rapidly dominates over the Yukawa force due to the finite
plasma wavelength of the metal.

\subsubsection{Magnetic and Other Backgrounds} {\bf{Magnetic
Susceptibility.}} Local magnetic field gradients can be caused by
the variation in magnetic susceptibility of the two source mass
materials.  For a background magnetic field, e.g. from the Earth,
the induced magnetic dipole moment in the materials produces a
field which varies in proximity to the two materials.  The induced
magnetization in a paramagnetic or diamagnetic material satisfies
\[
\vec{M} = \frac{\chi_m}{(1+\chi_m)\mu_0} \vec{B}
\]
Now the induced field due to the magnetized materials we roughly
approximate using the expression for a magnetized sphere
\[
B(z) = \frac{2 \mu_0 M a^3}{3z^3} = \frac{2 \chi_m}{(1+\chi_m)}
{B_0} \frac{a^3}{3z^3}
\]
where $a$ is the sphere radius and $\vec{B_0}$ is the background
magnetic field responsible for the induced magnetization.  We find
the differential field to be
\begin{eqnarray*} \Delta B_z (z) &=&
\frac{2 \Delta \chi}{(1+\chi)} B_0 \frac{a^3}{z^3} (\frac{\Delta z}{z}) \\
&\sim& \frac{2}{100} \Delta \chi B
\end{eqnarray*}
where in the last line we assume $\Delta z = 420$ nm and we take
the radius to be $a = 50\mu$m.  Silver and Gold are both
diamagnetic, with a differential susceptibility of $.6 \times
10^{-5}$.  We estimate $\Delta B \sim 10^{-7} B$.  The frequency
shift due to magnetic fields for Rubidium is $1.4$ MHz/Gauss, so
to obtain $10^{-4}$ Hz resolution only requires a magnetic field
shielding of $B_0 < 1$ mG.  One can further alleviate the
constraints on background magnetic fields by choosing materials
that have more similar magnetic susceptibility either in pure form
or through selective doping. Although not problematic for the
initial proposed parameters, extending the Yukawa sensitivity down
to the $\sim 10^{-7}$ Hz will require more extensive magnetic
shielding or precisely tailored alloy or doped materials.

{\bf{Gravity as a background.}} Although heavy nearby objects can
be easily detected, this is not expected to be problematic since
they in general cannot exhibit the periodicity of the source mass
pattern.  We note that in order to avoid acquiring a differential
background signal at the $10^{-7}$ Hz level due to the
gravitational attraction of the proof masses themselves (which is
only power-law suppressed and so remains significant at distances
much greater than the $\lambda$ of interest below 1 $\mu$m), it is
necessary to limit the vertical extent of the proof masses to be
less than approximately $100 \mu$m.

Finally, we list a number of other systematic backgrounds which
are not expected to be problematic due to common-mode rejection.
They include patch field effects on the surface of the reflecting
metal shield, the roughness of the surface of the shield, the
background Earth's gravitational field.

We conclude this section with a table summarizing the expected
frequency shifts of selected systematics as they compare to the
Yukawa signal.  Improvements beyond the level we discuss may be
attainable by tailoring materials to have more similar
conductivity, magnetic susceptibility, and by going to low
temperatures.

\begin{table}
\caption[]{Summary of selected systematic effects.  The three
backgrounds: Newtonian, $\Delta$ Casimir, and magnetic are
evaluated for a surface separation, trap spacing, and shield
thickness of 420 nm.  Scaling these distances from 420 nm to 600
nm causes the differential Casimir signal to drop below the
$10^{-7}$ Hz level. } \vglue0.3cm
\begin{center}
\begin{tabular}{|c|c|}
\hline
Yukawa signal &  \\
$\lambda$ & $\Delta f$(Hz) \\
\hline \hline
$1 ~\mu $m  &  $1.2 \times 10^{-10} ~\alpha $  \\
$.6~ \mu $m  &  $3.6 \times 10^{-11} ~\alpha $  \\
$.3~ \mu $m  & $3.4 \times 10^{-12} ~\alpha $ \\
$.1~ \mu $m  & $1.8 \times 10^{-15} ~\alpha $  \\
 \hline
 Background & \\
 \hline
 Newton & $6.5 \times 10^{-9}$ Hz  \\
 Casimir & $8283$ Hz \\
 $\Delta \eta$ & $7 \times 10^{-11}$ \\
 $\Delta$Casimir & $5.7 \times 10^{-7}$ Hz \\
Magnetic & $3 \times 10^{-5}$ Hz $\times \frac{\Delta
\chi}{10^{-6}} B$ (mG)
\\ \hline
\end{tabular}
\end{center}
\end{table}
\section{Discussion}
The search for short-distance modifications to Newtonian gravity
has rapidly expanded over the past seven years.  We have
illustrated several possible examples of sub-millimeter physics
that may occupy a vast amount of still unexplored parameter space.
We have also described an experimental technique involving
interference from arrays of Bose-Einstein condensed atoms that
could extend the search by several orders of magnitude below a
micron.  Such techniques, if successful, allow access to an area
of phase space that is complementary to the super-micron reach of
upcoming torsion oscillator and micro-cantilever experiments and
could lead to exciting discoveries.

\section{appendix}

\begin{figure}
\includegraphics[width=1.0 \columnwidth]{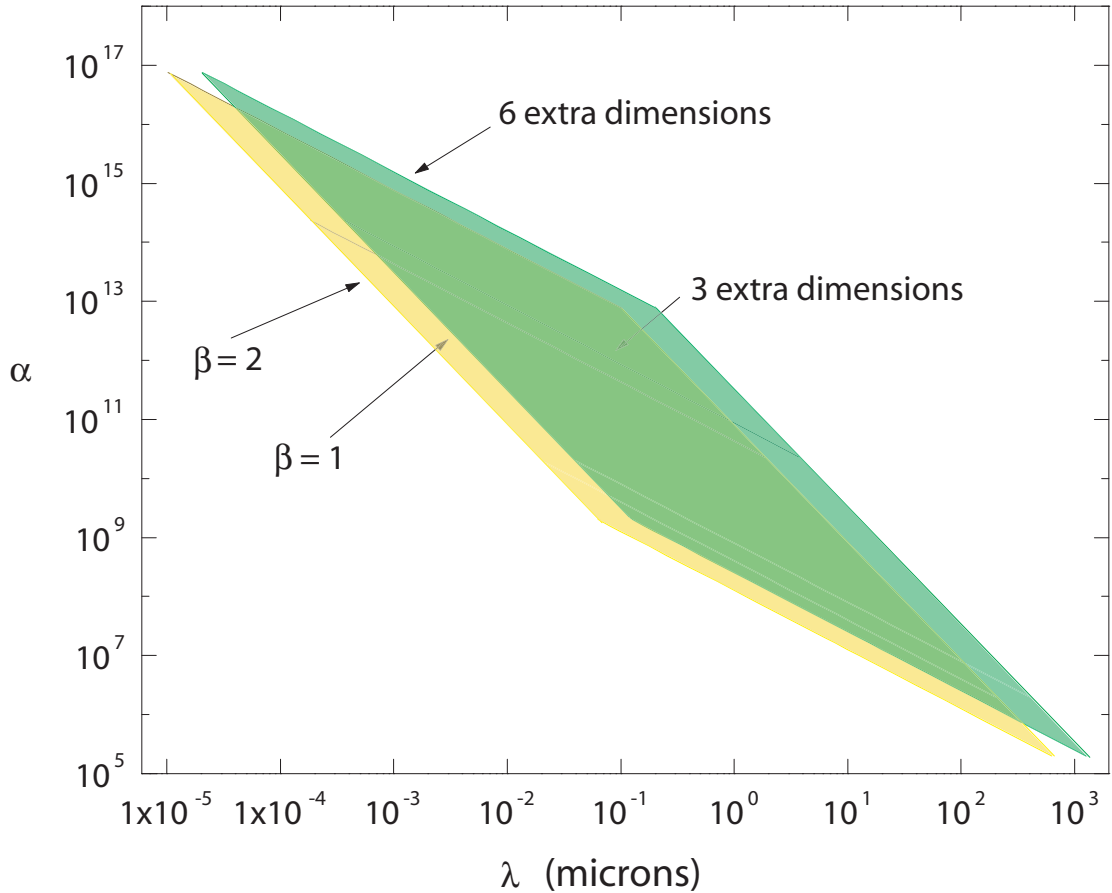}
\caption{Interaction strength and range for gauged Baryon number
in the bulk.  $\alpha = 1$ corresponds to a force of Newtonian
gravitational strength.  For illustration, the case of $\beta=1$
is shown as the green right-most parallelogram and $\beta=2$ is
shown in yellow.  The ranges for strong and weak coupling in 3 and
6 extra dimensions are shown.  The upper limit shown for the
strong-coupling region in the case of 3 extra dimensions
terminates at smaller alpha as shown.  The weak-coupling lower
boundary is identical for both cases.} \label{fig1}
\end{figure}

In this appendix we estimate the range and strength of forces due
to gauged baryon number in the bulk.  The mass of a macroscopic
object is roughly in proportion to the number of baryons, and the
expected ratio of the gauge to gravitational forces takes the form
\be \alpha = \frac{F_{gauge}}{F_{grav}} =
\frac{g^2_4}{4\pi}\frac{1}{G_N m^2_p} \ee where $m_p$ is the mass
of the proton. As discussed in the text, we take the 4+n
dimensional gauge coupling as $g^2_{4+n} = \Lambda^{-n} \O_{4+n}
\rho$, where $\Lambda$ is the ultraviolet cutoff scale and $\O_d =
\Omega_{(d-1)}/{(2\pi)^d} $ is the 1-loop suppression factor. If
the coefficient $\rho$ is ${\mathcal{O}}(1)$, this signifies
strong coupling at the scale $\Lambda$ in the 4+n dimensional
theory, as loop effects become comparable to tree-level. Using $
\Omega_{d-1} = 2\pi^{d/2}/\Gamma(\frac{d}{2})$, we find the
familiar 4-dimensional loop factor $\O_4 = 16 \pi^2$. For three
extra dimensions the factor becomes $\O_7 = 15 \cdot 8 \pi^4$ and
for six extra dimensions $\O_{10} = 12 \cdot 1024 \pi^5$. The
effective 4-d gauge coupling satisfies
\begin{eqnarray*}
\frac{1}{g_4^2} &=& \frac{V_d}{g^2_{(4+n)}}\\
                &=& (\frac{M_4}{M_*})^2 \frac{1}{V_{n-d}M_*^{n-d}} \frac{1}{\O_{d+4} \rho} (\frac{\Lambda^d}{M_*^d})
\end{eqnarray*}
where in the second line we have assumed that the gauge bosons
propagate in $d \leq n$ of the extra dimensions. For definiteness,
we consider the case of gauge bosons living in all extra
dimensions (n=d), and the expression simplifies to \be
g^2_4=(\frac{{M_*}^2}{{M_4}^2}) \O_{n+4} \cdot\rho
(\frac{M_*^n}{\Lambda^n}). \ee so then the ratio of forces becomes
\begin{eqnarray}
\alpha &=& \frac{g_4^2}{4\pi} \frac{1}{G_N m_p^2} \\
         &=& (\frac{{M_*}^2}{{m_p}^2}) 25~\frac{~\O_{n+4} \cdot\rho}{4\pi} (\frac{M_*^n}{\Lambda^n}).
         \label{alphaeq}
\end{eqnarray}
The factor of $25$ appears as the square of the approximate ratio
between $1/\sqrt{G_N}$ and the reduced Planck mass. For $M_*$ of
order TeV, already the force becomes $\sim$ 1 million times
gravity.  If the gauge symmetry is spontaneously broken by a
scalar field $\chi$ obtaining a vacuum expectation value
$\langle\chi\rangle$, the resulting mass of a gauge particle
$A_\mu$ becomes \be m_A = g_4 \langle \chi \rangle. \ee Here again
we assume that $\chi$ condenses on a brane other than our own.  It
was shown in reference \cite{dvali1} that forces mediated by bulk
gauge fields can be exponentially weaker than gravity if the bulk
gauge symmetry is spontaneously broken on our brane.

We parameterize the vacuum expectation value of $\chi$ as \be
<\chi>= \beta \cdot \Lambda \ee where beta is a numerical
coefficient of order 1, so that the Compton wavelength is written
\be \lambda_A = \frac{1}{\O_{n+4} \cdot \rho} \frac{M_4}{M_*^2}
\frac{1}{\beta} \frac{\Lambda^{n/2}}{M_*^{n/2}} \label{lameq} \ee
If the scale $\Lambda$ is somewhat smaller than $M_*$, the Compton
wavelength becomes shorter and the strength $\alpha_g$ increases.
For example, if one imagines new physics arising from string
theory at scale $M_s$, we have the relation
\[
\frac{M_s^{2+n}}{g_s^2} \sim M_*^{2+n}
\]
so that \be M_s = M_* (g_s^2)^{1/(2+n)} \leq M_*. \ee where we
have identified the Lagrangian \be \int{d^4 x} \sqrt{-g}
{\mathcal{R}} M_*^{2+n} V_n = \int{d^4 x} \sqrt{-g} {\mathcal{R}}
M_4^{2} \ee with a Type I string Lagrangian \cite{iadd}
\[
\int{d^4 x \frac{1}{g_s^2} M_s^{2+n} V_n \sqrt{-g} {\mathcal{R}}}.
\]
If $\Lambda$ is associated with physics at the string scale, it is
possible the coefficient $\beta$ is greater than 1.

In order to illustrate the phase space encompassed by these
forces, in Fig. 1 we plot $\alpha$ versus $\lambda_A$, where for
simplicity we set $\Lambda = M_*$ and we vary the parameter $\rho$
from $1/(10 \times \O_{d+4})$ for weak coupling, to 1 for strong
coupling.  We vary $M_*$ from 1 TeV to 100 TeV, and show $\beta$
between 1 and 2.  For smaller $\beta$, the predicted region can be
extended horizontally to the right until conflicting with
experimental observation.

Finally we note the gauge force we discuss is due strictly to the
zero-mode, and therefore its range is not strongly limited by the
size of the extra dimensions.  In the case that any of the extra
dimensions has a large enough compactification radius, the
lightest KK modes of the gauge particles may also make a
contribution, though we do not include this explicitly in Figures
1 and 2.

\begin{acknowledgments}
We are indebted to Mark Kasevich who contributed immensely to the
development of this work.  We also thank Ignatios Antoniadis, Blas
Cabrera, Eric Cornell, Chad Davis, Peter Graham, and Vladan
Vuletic for useful discussions. This work is supported by grant
NSF-PHY-9870115.
\end{acknowledgments}

\def\ijmp#1#2#3{{\it Int. Jour. Mod. Phys. }{\bf #1~}(19#2)~#3}
\def\pl#1#2#3{{\it Phys. Lett. }{\bf B#1~}(19#2)~#3}
\def\zp#1#2#3{{\it Z. Phys. }{\bf C#1~}(19#2)~#3}
\def\prl#1#2#3{{\it Phys. Rev. Lett. }{\bf #1~}(19#2)~#3}
\def\rmp#1#2#3{{\it Rev. Mod. Phys. }{\bf #1~}(19#2)~#3}
\def\prep#1#2#3{{\it Phys. Rep. }{\bf #1~}(19#2)~#3}
\def\pr#1#2#3{{\it Phys. Rev. }{\bf D#1~}(19#2)~#3}
\def\np#1#2#3{{\it Nucl. Phys. }{\bf B#1~}(19#2)~#3}
\def\mpl#1#2#3{{\it Mod. Phys. Lett. }{\bf #1~}(19#2)~#3}
\def\arnps#1#2#3{{\it Annu. Rev. Nucl. Part. Sci. }{\bf
#1~}(19#2)~#3}
\def\sjnp#1#2#3{{\it Sov. J. Nucl. Phys. }{\bf #1~}(19#2)~#3}
\def\jetp#1#2#3{{\it JETP Lett. }{\bf #1~}(19#2)~#3}
\def\app#1#2#3{{\it Acta Phys. Polon. }{\bf #1~}(19#2)~#3}
\def\rnc#1#2#3{{\it Riv. Nuovo Cim. }{\bf #1~}(19#2)~#3}
\def\ap#1#2#3{{\it Ann. Phys. }{\bf #1~}(19#2)~#3}
\def\ptp#1#2#3{{\it Prog. Theor. Phys. }{\bf #1~}(19#2)~#3}

\bibliography{}

\begin{thebibliography}{00}
\bibitem{sg}
S. Dimopoulos and G.F.Giudice, Phys.Lett. {\bf{B379}}, 105-114
(1996).
\bibitem{add97} I. Antoniadis, S. Dimopoulos, G. Dvali
Nucl.Phys. {\bf{B516}}, 70-82 (1998).

\bibitem{add1}
Nima Arkani-Hamed, Savas Dimopoulos, and Gia Dvali. Phys.Lett.
{\bf{B429}}, 263-272 (1998).
%
\bibitem{iadd}
Ignatios Antoniadis, Nima Arkani-Hamed, Savas Dimopoulos, Gia
Dvali, Phys.Lett. {\bf{B436}} ,257-263 (1998) ,hep-ph/9804398.
%
\bibitem{add3}
Nima Arkani-Hamed, Savas Dimopoulos, and Gia Dvali. Phys.Rev.
{\bf{D59}} ,086004 (1999), hep-ph/9807344.
%
\bibitem{raman97} T. Banks, Nucl.Phys.{\bf{B 309}}, 493 (1988). \\ S.R.Beane, Gen.Rel.Grav.{\bf{29}}, 945-951 (1997).
\\ Raman Sundrum,  JHEP 9907:001,(1999).
%

\bibitem{price} Joshua C. Long, Allison B. Churnside, John C.
Price, Proceedings of the Ninth Marcel Grossmann Conference (Rome,
2-8 July 2000), hep-ph/0009062.
\bibitem{adelberger}  C. D. Hoyle, U. Schmidt, B. R. Heckel, E. G. Adelberger, J. H. Gundlach, D. J. Kapner, H. E. Swanson,
 Phys.Rev.Lett. {\bf{86}} , 1418-1421 (2001) .
\bibitem{stanford} J. Chiaverini, S. J. Smullin, A. A. Geraci, D. M. Weld, A. Kapitulnik
, Phys.Rev.Lett. {\bf{90}}, 151101 (2003).
\bibitem{pricenature} Joshua C. Long, Hilton W. Chan, Allison B. Churnside, Eric A. Gulbis, Michael C. M. Varney, John C. Price
, Nature 421, 922 - 925 (2003).

\bibitem{pricereview}  For a recent review see Joshua C. Long and John C. Price, to be published in Comptes Rendus Physique
hep-ph/0303057.

\bibitem{giudicereview}
G. F. Giudice and R. Rattazzi, Phys.Rept. {\bf{322}}, 419-499
(1999),
 hep-ph/9801271.
\bibitem{fischbach}  E. Fischbach, D. E. Krause, V. M. Mostepanenko, M.
Novello, Phys.Rev. {\bf{D64}} , 075010 (2001).
%
\bibitem{irvine} J.K.~Hoskins {\it et al.}, \pr{32}{85}{3084}.

\bibitem{lamoreaux} S.K. Lamoreaux, Phys.Rev.Lett. {\bf{78}},5
(1997).

\bibitem{ven} G.~Veneziano and T.~Taylor, \pl{213}{88}{450}.

\bibitem{kw}  David B. Kaplan, Mark B. Wise, JHEP 0008, 037
(2000).
%
\bibitem{dp} T. Damour, A. M. Polyakov,  Nucl.Phys. {\bf{B423}},
532-558 (1994).
%
\bibitem{kands}
A. Kehagias and K. Sfetsos, Phys.Lett. {\bf{B472}}, 39-44 (2000) ,
hep-ph/9905417.
%
\bibitem{fandl}
E.G.Floratos and G.K. Leontaris, Phys.Lett. {\bf{B465}}, 95-100
(1999), hep-ph/9906238.
%
\bibitem{raffelt} Steen Hannestad and Georg G. Raffelt, Phys.Rev.Lett.
{\bf{88}} ,071301 (2002).

\bibitem{raffelt2} S.Hannestad, G.G.Raffelt, Phys.Rev. {\bf{D67}}
125008(2003).

\bibitem{dgkn} G. Dvali, G. Gabadadze, M. Kolanovic, and F. Nitti, Phys.Rev. {\bf{D64}} 084004 (2001).

\bibitem{nimasavas}
Nima Arkani-Hamed and Savas Dimopoulos,Phys.Rev. {\bf{D65}},
052003 (2002), hep-ph/9811353.

\bibitem{dvali1}
Gia Dvali, Gregory Gabadadze, Massimo Porrati, Mod.Phys.Lett.
{\bf{A15}}, 1717-1726 (2000), hep-ph/0007211.

\bibitem{cullenmaxim}
Schuyler Cullen, Maxim Perelstein,  Phys.Rev.Lett. {\bf{83}},
268-271 (1999).
\bibitem{nobel1} S.Chu, S.Cohen-Tannoudji, and W.Phillips, Nobel
lectures (1998).

\bibitem{pcc} A. Peters, K.Y. Chung, and S. Chu, Metrologia,
{\bf{38}}, 25-61, 2001.

\bibitem{mffsk} J.M. McGuirk, G.T. Foster, J.B.Fixler,
M.J.Snadden, and M.A.Kasevich, Phys.Rev. {\bf{A65}} 033608 (2002).

\bibitem{nobel2} E.Cornell, W.Ketterle, and C.Wieman, Nobel lectures
2001.
\bibitem{leggett} Anthony J. Leggett, Rev.Mod.Phys.{\bf{73}},307(2001)
\bibitem{andersonkasevich98} B.P.Anderson and M.A.Kasevich,
Science 282, 1686 (1998).
\bibitem{mark} Mark Kasevich, private communication.

\bibitem{feshbach}
E. Tiesinga et al, Phys. Rev. {\bf{A46}}, R1167 (1992).
\\ E. Tiesinga,
B.J.Verhaar, and H.T.C.Stoof, Phys. Rev. {\bf{A47}} 4114 (1993).\\
J.L.Roberts, N.R. Claussen, James P. Burke, Chris H. Greene,
E.A.Cornell, and C.E.Wieman, Phys. Rev. Lett. {\bf{81}}, 5109 (1998). \\
S.L.Cornish, N.R. Claussen, J.L.Roberts, E.A.Cornell, and C.E.
Wieman, Phys. Rev. Lett. {\bf{85}}, 1795 (2000).
\bibitem{thermal}
D.S. Petrov, G.V. Shlyapnikov, and J.T.M. Walraven, Phys. Rev.
Lett. {\bf{87}}, 050404 (2001). \\ S. Dettmer, D. Hellweg, P.
Ryytty, et al., Phys. Rev. Lett. {\bf{87}}, 160406 (2001).
\bibitem{surface}
D.M. Harber, J.M. McGuirk, J.M. Obrecht, E.A. Cornell, JLTP in
press, cond-mat/0307546.\\ Bradley J. Roth, J. Appl. Phys.
{\bf{83}} 635 (1998). \\ C. Henkel, S. Potting, M. Wilkens, Appl.
Phys. B 69, 379 (1999). \\  A.E. Leanhardt, Y. Shin, A.P.
Chikkatur, D. Kielpinski, W. Ketterle, D.E. Pritchard, Phys. Rev.
Lett. {\bf{90}}, 100404 (2003). \\ M. P. A. Jones, C. J. Vale, D.
Sahagun, B. V. Hall, and E. A. Hinds, Phys. Rev. Lett. {\bf{91}},
080401 (2003).
\bibitem{lambrecht} Astrid Lambrecht, Serge Reynaud,
Eur.Phys.J. D8, 309-318 (2000) quant-ph/9907105.
\bibitem{lifshitz} L.D.Landau and E.M. Lifshitz. Statistical
Physics vol. 2, ch 13.
\bibitem{panenko} B. Geyer, G.L. Klimchitskaya, V.M.Mostepanenko
Phys. Rev. {\bf{A65}}, 062109 (2002).


\end{thebibliography}


\end{document}